\documentclass[%
reprint,
superscriptaddress,
bibnotes,
amsmath,amssymb,
aps,
prb,
onecolumn,
]{revtex4-2}

\usepackage[main=english,ngerman]{babel}
\usepackage[T1]{fontenc}
\usepackage[utf8]{inputenc}
\usepackage{lmodern}
\usepackage{microtype}
\usepackage{graphicx}
\usepackage{hyperref}
\usepackage{xr-hyper}
\usepackage{amsfonts}
\usepackage{siunitx}
\usepackage{physics}
\usepackage{mathtools}
\usepackage{cleveref}
\usepackage{xfrac}
\usepackage{placeins}
\makeatletter
\g@addto@macro\bfseries{\boldmath}
\makeatother

\newcommand{\PP}{\mathrm{P}} % Polariton
\newcommand{\MM}{\mathrm{M}} % Magnon
\newcommand{\Pol}{\mathrm{Pol}} % Polariton (long form)
\renewcommand{\qq}{\mathbf{q}} % Wave vector q (overrides \qq from the physics package)
\newcommand{\kk}{\mathbf{k}} % Wave vector k

\begin{document}

\title{Magnon--Phonon--Polariton Optomechanics in a van der Waals Magnet}

\author{Jens-Christian Drawer}%
\affiliation{Institut für Physik, Fakultät V, Carl von Ossietzky Universität Oldenburg, 26129 Oldenburg, Germany}%

\author{Vladislav Yu.\ Shishkov}%
\affiliation{Macroscopic Quantum Optics (MQO) Labs, Department of Applied Physics, Aalto University School of Science, FI-00076 Espoo, Finland}%

\author{Bo Han}%
\affiliation{Institut für Physik, Fakultät V, Carl von Ossietzky Universität Oldenburg, 26129 Oldenburg, Germany}%

\author{Vita Solovyeva}%
\affiliation{Institut für Physik, Fakultät V, Carl von Ossietzky Universität Oldenburg, 26129 Oldenburg, Germany}%

\author{Iva Plutnarová}%
\affiliation{Department of Inorganic Chemistry, Faculty of Chemical Technology, University of Chemistry and Technology Prague, Technick\'{a} 5, Prague 6, 16628, Czech Republic}%

\author{Zdeněk Sofer}%
\affiliation{Department of Inorganic Chemistry, Faculty of Chemical Technology, University of Chemistry and Technology Prague, Technick\'{a} 5, Prague 6, 16628, Czech Republic}%

\author{Christian Schneider}%
\affiliation{Institut für Physik, Fakultät V, Carl von Ossietzky Universität Oldenburg, 26129 Oldenburg, Germany}%

\author{Anton V.\ Zasedatelev}%
\affiliation{Macroscopic Quantum Optics (MQO) Labs, Department of Applied Physics, Aalto University School of Science, FI-00076 Espoo, Finland}%

\author{Martin Esmann}%
\thanks{Corresponding author: \href{mailto:m.esmann@uni-oldenburg.de}{m.esmann@uni-oldenburg.de}}
\affiliation{Institut für Physik, Fakultät V, Carl von Ossietzky Universität Oldenburg, 26129 Oldenburg, Germany}%

\date{\today}%

\begin{abstract}
	The layered anti-ferromagnet CrSBr hosts self-hybridized exciton--polaritons that couple parametrically to coherent magnons and phonons. In this work, we explore the tripartite optomechanical interaction between these quasiparticles in suspended CrSBr membranes supporting high-$Q$ breathing-type acoustic phonon modes. Using resonant pump--probe spectroscopy on the polariton state, we resolve the coupled dynamics in the $\simeq\SI{20}{\GHz}$ range while sweeping the magnon mode across the phonon resonance with an external magnetic field.
	We observe an attractive magnon--phonon interaction of strength $V/2\pi = \SI{-43}{\MHz}$, which primarily leads to phonon amplification. At the magnon-phonon resonance, we find predominantly unidirectional energy flow from magnons to phonons while the possible back-action is limited by kinetic losses of coherent magnons. Our study enables direct interfacing of collective magnetic excitations with sound and light in layered van der Waals materials, paving the way for quantum-coherent transduction in the \unit{\GHz} range with optical readout.
\end{abstract}

\maketitle

\section{Introduction}

Hybrid optomechanical quantum systems combine the spectral selectivity and geometric tunability of sharp mechanical resonances with the sensitive, non-contact excitation and readout provided by optics~\cite{chu2017,delsing2019,clerk2020}. Another class of hybrid quantum systems leverages the coupling of mechanical resonances to collective spin excitations. Magnons, the quanta of spin waves, can transport information without charge flow. Coupling them to mechanical motion may be used to extend coherent magnon propagation length~\cite{an2020,ruckriegel2020}. Coupling magnons to mechanical excitations has been investigated in nanomagnets, low-dimensional devices~\cite{vaclavkova2021,awschalom2021,hwang2024,kim2025}, and established platforms such as yttrium iron garnet (YIG)~\cite{zhang2016,zhang2023a}. 
In YIG-based cavity magnomechanics, magnons couple to phonons via magnetostriction and to microwave photons via the magnetic-dipole interaction~\cite{zhang2016,zuo2024a}, while optomagnonics extends this coupling to optical photons through magneto-optical effects~\cite{violakusminskiy2016}. For such tripartite systems, theory predicts magnon--photon--phonon entanglement~\cite{li2018}, mechanically mediated magnon-to-optics quantum state transfer~\cite{fan2023}, microwave--optics entanglement~\cite{fan2023a}, and broadband magnomechanical frequency conversion~\cite{engelhardt2022}, and magnon-mediated microwave-to-optical conversion~\cite{hisatomi2016}. However, in these platforms, the magnetic, mechanical, and optical degrees of freedom reside in separate components or couple only weakly, which limits conversion efficiencies~\cite{hisatomi2016,zuo2024a}.
This motivates the search for materials and hybrid quantum architectures in which coupled optical, mechanical, and magnetic degrees of freedom coexist within strongly coupled light--matter states, such as polaritons, enabling quantum coherent \unit{\GHz} transduction with optical readout.

Van der Waals (vdW) magnets are well suited to this setting~\cite{gibertini2019a}, with the ambient-stable semiconductor CrSBr being a particularly promising example. CrSBr is an A-type antiferromagnet below $T_\mathrm{N}\simeq\SI{132}{\K}$~\cite{lee2021a,wilson2021a,rizzo2022a,ye2022a}. The excitonic transitions in CrSBr are sensitive to magnetic order and interlayer electronic hybridization, enabling optical access to magnetic order~\cite{wilson2021a,dirnberger2023} and coherent magnons~\cite{bae2022,diederich2023,bae2024,komar2024a}. The magnetic state can additionally be modified by strain and pressure~\cite{cenker2022a,pawbake2023,cenker2025}, electrostatic doping~\cite{tabataba-vakili2024a,hong2025a}, magnetic fields~\cite{wilson2021a}, and twist~\cite{mondal2025a}. Strong exciton--photon coupling further produces self-hybridized exciton--polaritons in CrSBr slabs ~\cite{dirnberger2023,wang2023c} and optical cavities~\cite{wang2023a,ruta2023b,li2024a}. Self-hybridized polariton resonances are therefore a straightforward way to combine optical accessibility with the magnetic and strain sensitivity of the exciton in a hybrid quantum system.

Suspending vdW membranes provides a natural means to engineer their mechanical spectrum into a well-defined comb of high $Q$ acoustic phonon modes~\cite{soubelet2019,zalalutdinov2021,yoon2024,carr2024,aversa2026}. In contrast to magnon--phonon coupling scenarios~\cite{zhang2023a} based on material-intrinsic optical phonon resonances~\cite{mai2021}, surface acoustic waves~\cite{lyons2023}, nanopatterning~\cite{berk2019,kim2025}, or macroscopic deformation modes~\cite{zhang2016}, a suspended membrane supports thickness-defined, breathing-type Fabry--Pérot modes whose frequencies can be tailored geometrically. Magnetoelastic coupling to such discrete acoustic eigenmodes has been treated theoretically for thin films~\cite{streib2018}, finite magnets~\cite{gonzalez-ballestero2020,gonzalez-ballestero2020a}, and suspended vdW membranes~\cite{das2026}.

This approach is particularly useful for magnon--phonon coupling because an external magnetic field tunes the magnon resonance, allowing it to be brought into mutual resonance with a selected mechanical mode. While coherent magnon--phonon dynamics based on intrinsic phonon resonances~\cite{shen2025} and coupled magnon--polariton dynamics~\cite{dirnberger2023} are established in CrSBr, its large potential for simultaneous magnon--polariton, phonon--polariton, and magnon--phonon coupling based on breathing-type resonances has not been explored.

Here, we investigate tripartite polariton--magnon--phonon dynamics in suspended CrSBr membranes with engineered acoustic breathing modes. Resonant pump--probe spectroscopy of the polariton~\cite{thomsen1984,thomsen1986,ruello2015} provides optical access to coherent dynamics near $\simeq\SI{20}{\GHz}$, while an external magnetic field tunes the magnon through the mechanical resonance. We observe an attractive magnon--phonon interaction of strength $V/2\pi = \SI{-43}{\MHz}$, which leads to phonon amplification at the resonance. The energy transfer is largely unidirectional, while phonon back-action is limited only by kinetic losses of the coherent magnons. These results establish optically detected tripartite polariton--magnon--phonon dynamics in a layered antiferromagnet and demonstrate engineered acoustic resonances coupled to coherent spin excitations with optical readout.

\section{Methods and results}

\begin{figure}[ht]
	\includegraphics{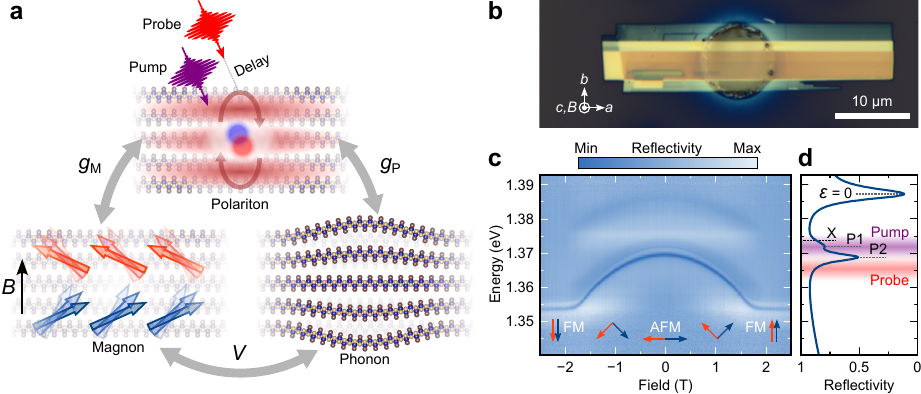}%
	\caption{\label{fig:1}%
		(a)~Schematic representation of the tripartite coupling scenario in a suspended CrSBr slab. Self-hybridized exciton--polaritons (top) parametrically couple to coherent magnons (bottom left) and breathing-type phonons (bottom right) around \SI{20}{\GHz}. The latter two exchange energy via a direct magnon--phonon coupling $V$. The pump--probe setup (top) resonantly generates exciton--polaritons and detects the tripartite dynamics via changes in transient probe reflectivity $\sfrac{\Delta R(t)}{R_0}$. (b)~Optical microscope image of a CrSBr slab suspended over a hole, with the crystal axes and applied magnetic field orientation as indicated. (c)~Measured optical reflection spectrum of the CrSBr slab showing exciton--polariton resonances that red-detune as a function of applied magnetic field. The bottom schematic illustrates the spin-canting configuration with spins aligned at $B_\mathrm{sat}\approx \SI{1.6}{\T}$. (d)~Simulated optical reflectivity of the CrSBr slab calculated by transfer matrix method and slab thickness obtained from AFM measurement. P1 and P2 mark the lower exciton--polariton resonances, X the position of the bare exciton and $\varepsilon=0$ the resonance for which the real part of the CrSBr dielectric function dips through zero. The pump and probe pulses are tuned to the flanks of the exciton--polariton resonance.}%
\end{figure}

\Cref{fig:1}(a) outlines the polariton-optomechanical coupling scenario explored in this work: A thin slab of the layered anti-ferromagnet CrSBr hosts tightly bound excitons, which couple strongly to confined photon modes in the slab, forming self-hybridized exciton--polaritons (top). When the slab is mechanically suspended, out-of-plane acoustic (LA) phonons along the $\Gamma$--Z direction in the vdW slab yield high $Q$ phonon modes with resonances in the \qtyrange{10}{100}{\GHz} range (bottom right), which parametrically couple to the polariton states via a deformation--potential interaction $g_\PP$, i.e., by periodically strain-tuning the exciton--polariton energy. Similarly, CrSBr exhibits out-of-plane magnons in the \SI{20}{\GHz} range which interact with excitons (bottom left), resulting in a second parametric mechanism of coupling to the exciton--polaritons $g_\MM$. Magnetic field-tuning of the magnon frequency and engineering the phonon mode spectrum via the slab thickness enable us to explore these two parametric couplings independently, while bringing the magnon and phonon into mutual resonance leads to an additional direct interaction $V$ between them. We coherently excite and probe the system's optomechanical dynamics by two-color pump--probe spectroscopy, resonantly addressing an exciton--polariton resonance (\cref{fig:1}(a), top). While the pump pulse quasi-instantaneously generates a polariton population, the probe detects transient changes in reflectivity $\sfrac{\Delta R(t)}{R_0}$. \Cref{fig:1}(b) shows an optical image of our sample. By dry-stamping, we suspended an up to \SI{75}{\nm} thick CrSBr slab over a \SI{10}{\um}-diameter \SI{1}{\um}-deep hole prepared in a Si/SiO\textsubscript{2} substrate by focused ion--beam (FIB) milling. The external magnetic field tuning the magnon frequency is applied out-of-plane, along the crystal $c$-axis. The slab contains multiple steps, visible as color contrasts, that correspond to sections with different thickness, i.e., acoustic Fabry--Pérot modes with different breathing-type resonant frequencies. \Cref{fig:1}(c, left) shows a low-temperature reflectivity spectrum of the sample in panel~(b) as a function of the external magnetic field strength. With the bare exciton at \SI{1.374}{\eV}, we observe two close-by self-hybridized lower polariton modes, and a fainter reflectivity dip at \SI{1.387}{\eV}, above the exciton. All three resonances red-shift with external field up to a maximum shift of \SI{17}{\meV} for a saturation field strength of $B_\mathrm{sat}=\SI{1.6}{\T}$. This commonly observed shift has been explained as the consequence of forcing the anti-ferromagnetic spin-order into out-of-plane alignment (see sketch at the bottom of panel~(c)), which in turn leads to delocalization of the exciton over multiple vdW layers and a reduction in exciton binding energy~\cite{wilson2021a}. This mechanism is responsible for the pronounced parametric coupling between self-hybridized polariton resonances and the out-of-plane magnon in CrSBr. \Cref{fig:1}(d) shows a simulation of the optical reflectivity spectrum at $B=0$ obtained with a transfer matrix method using the dielectric function of CrSBr along its $b$-axis reported in~\cite{dirnberger2023}, together with the slab thickness measured by AFM. X marks the spectral position of the uncoupled exciton, whereas P1 and P2 are the two lower polariton resonances also observed in the measurement. By comparison to the dielectric function, we assign the additional resonance observed in the measurement to the position for which the real part dips through zero, i.e., $\varepsilon=0$. We spectrally align the pump and probe laser on the two flanks of polariton resonance P2 as indicated by the colored bars in panel~(d), where the reflectivity is most sensitive to spectral shifts.

To access the tri-partite magnon--phonon--polariton interaction depicted in \cref{fig:1}(a), we perform local pump--probe spectroscopy at a temperature of \SI{4}{\K}. For a detailed description of the optical setup, refer to \cref{SI-sec:ExperimentalMethods} of the Supporting Information. We use a Ti:Sapphire laser emitting pulses with \SI{140}{\fs} nominal duration at a repetition rate of \SI{80}{\MHz} and a central wavelength of \SI{905}{\nm}. We prepare two orthogonally polarized pulse trains, and send one through a motorized mechanical delay line. The relative power of the two pulses in each pair is controlled with a $\sfrac{\lambda}{2}$-waveplate in front of a polarizing beam splitter (PBS). We derive separate colors for pump and probe by first introducing a shortpass (SP) filter into the pump path. The two beams are then recombined in a second PBS and focused tightly onto the sample, to an estimated focal diameter of \SI{1}{\um}. In this cross-polarization scheme, the first pulse with an average power of \SI{1}{\mW} acts as the pump, coherently driving exciton--polaritons in the CrSBr membrane. Parametric coupling entails the generation of coherent magnon and phonon excitations through $g_\MM$ and $g_\PP$, respectively. The second pulse arriving on the sample at a variable delay $t$, containing the full laser spectrum with an average power of \SI{0.1}{\mW}, then acts as a probe polarized nearly parallel to the $b$-axis and detects transient changes in optical reflectivity $\sfrac{\Delta R(t)}{R_0}$ caused by strain of the phonon wave packet and the magnon-induced modulation of the spin canting angle, both of which shift the polariton resonance. The reflected light is passed through a set of waveplates, a polarizer, and a longpass (LP) filter with its transmission edge at \SI{910}{\nm}, which together suppress the pump before the reflected probe reaches a fast photodetector. The spectral filtering leads to effective pulse durations for both pump and probe of \SI{400}{\fs}. We modulate the pump at \SI{1}{\MHz} by an acousto-optic modulator and detect pump-induced changes in reflectivity via lock-in demodulation of the photodetector signal.

\begin{figure}[ht]
	\includegraphics{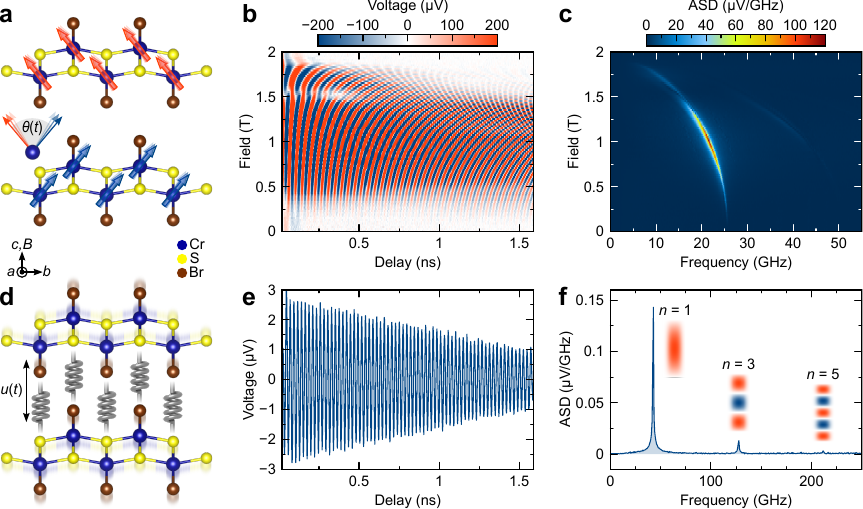}%
	\caption{\label{fig:2}%
		(a)~Schematic illustration of the spin dynamics in the crystal under an applied magnetic field. In neighboring layers, the spins precess in phase about their mean, field-dependent equilibrium tilt angle, such that the enclosed angle $\theta$ varies periodically in time. (b)~Pump--probe time traces acquired on the CrSBr slab in \cref{fig:1}(b) next to the suspended region, showing the magnon-induced periodic modulation of the sample's transient reflectivity. (c)~Amplitude spectrum of the first time derivative of the signal shown in (b). (d)~Schematic illustration of the breathing-type LA phonon mode in CrSBr with a relative displacement between neighboring layers $u(t)$ strain-coupled to the exciton--polariton resonance via a deformation potential mechanism. (e)~Pump--probe time trace showing the acoustic-phonon-induced modulation of the transient reflectivity of the sample for zero magnetic field. (f)~Amplitude spectrum of the first time derivative of the signal shown in (e). The strain distributions across the crystal corresponding to the observed resonances are schematically illustrated.}%
\end{figure}

In \cref{fig:2}(a--c), we show the coherent generation and detection of magnons by locally performing pump--probe spectroscopy on a part of the sample in \cref{fig:1} that is not mechanically suspended. LA phonons are strongly damped by propagation into the Si/SiO\textsubscript{2} substrate, and therefore, no mechanical Fabry--Pérot resonances are observed. When an external magnetic field is applied with a flux density $B<B_\mathrm{sat}$, the spin order changes from anti-ferromagnetic to a canted arrangement, as sketched in panel~(a). An out-of-plane magnon mode thus corresponds to a harmonic modulation of the canting angle $\theta$ and hence the net magnetization along the $c$-axis. Panel~(b) shows magnetic field-dependent pump--probe traces over pump--probe delays up to $t=\SI{1.6}{\ns}$ with the full trace extending up to $t=\SI{11}{\ns}$, i.e., the maximum possible delay between subsequent pump pulses. In agreement with previous reports~\cite{bae2022,dirnberger2023}, we observe pronounced coherent oscillations for fields in the range $\SI{100}{\mT}<B<\SI{1.9}{\T}$, with the oscillation frequency reducing as the field is ramped up. In the field-free and the saturated regime, these oscillations are absent. Interestingly, the saturation field appears to be slightly larger on the supported CrSBr part than for the suspended area probed in \cref{fig:1}(c), potentially due to static strain. By performing a Fourier transformation for each trace at fixed field strength $B$, we extract the field-tunable magnon frequency. In agreement with the literature~\cite{bae2022,dirnberger2023}, the fundamental out-of-plane magnon frequency tunes from \SI{25}{\GHz} for external fields in the \SI{100}{\mT} range towards zero when approaching field saturation. Its quality factor in the range of \SI{1}{\T} is on the order of \num{100}. We also observe a much fainter contribution from the second harmonic of the magnon at \SI{40}{\GHz} for intermediate field strengths around \SI{1}{\T}.

In \cref{fig:2}(d--f), we show the coherent generation and detection of breathing-type acoustic phonon modes in the suspended part of the CrSBr slab. In this measurement, the external magnetic field is switched off, such that no coherent magnons are generated. As sketched in panel~(d), the coherent phonon pulse can be thought of as the out-of-plane motion of adjacent CrSBr layers, each representing a mass density connected via a vdW spring exerting a restoring force. The associated strain, $\dv{u}{z}$ with $u(z)$ the mechanical displacement, couples the phonon to the polariton resonance via a deformation potential mechanism~\cite{soubelet2019, zalalutdinov2021, yoon2024,carr2024,aversa2026}. In panel~(e), we plot a pump--probe trace at $B=\SI{0}{\T}$ for the same range of pump--probe delays up to $t=\SI{1.6}{\ns}$. We observe clear oscillations with a decaying envelope, representing the phonon amplitude lifetime. By fitting an exponentially decaying sinusoid to the trace, we obtain a quality factor of \num{215.8\pm2.1} for its dominating \SI{43}{\GHz} component. A Fourier transform of this trace yields the amplitude spectrum in \cref{fig:2}(f). We see pronounced resonances at \SI{43}{\GHz}, \SI{128}{\GHz}, and \SI{210}{\GHz}, which correspond to the first, third, and fifth order breathing-type phonon modes, with their strain patterns sketched as insets. Due to vanishing overlap integrals of the optical intensity pattern of the excitation pulse and the strain distribution of the even harmonics, only the odd harmonics appear in this spectrum~\cite{fainstein2013,lanzillotti-kimura2011}. Unlike the magnon resonance, the frequency comb of acoustic phonon breathing modes is independent of the applied magnetic field. However, this frequency comb can be engineered by choosing a specific thickness of the suspended CrSBr slab. Essentially, these breathing-type modes are Fabry--Pérot resonances. With $v_g$ the speed of sound for out-of-plane acoustic phonons in CrSBr, the $m$-th harmonic in a slab of thickness $d$ appears at a frequency of $f_m=m\cdot\frac{v_g}{2d}$. This spectrum is limited at the top by the acoustic phonon frequency at the edge of the first Brillouin zone.

\begin{figure}[ht]
	\includegraphics{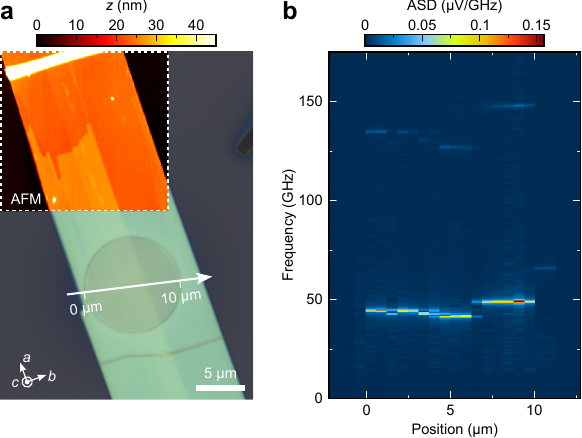}%
	\caption{\label{fig:3}%
		(a)~Optical microscope image of a CrSBr slab suspended over a hole. The thickness of the slab is measured in an atomic force microscopy (AFM) scan, and the $z$-profile is indicated as a false color map in the white dashed box overlaid on the image. A spatially resolved pump--probe measurement in the absence of a magnetic field is performed step wise along a line across the hole, as indicated by the white arrow. (b)~Amplitude spectra of the first time derivative of the signals obtained along the path across the drum indicated in (a). The breathing-type phonon resonances are thickness-dependent enabling nanophononic engineering of the tripartite coupling in CrSBr.}%
\end{figure}

To further demonstrate this thickness-dependent phonon engineering, we performed local pump--probe spectroscopy on various parts of a suspended CrSBr membrane with different number of vdW layers. The results are compiled in \cref{fig:3}. Panel~(a) shows an optical image of the suspended slab (green), with the inset in the dashed white box showing an AFM image from the indicated region. We determine the height of each section on the slab in the range of \SI{20}{\nm}. We measured pump--probe traces along the white line drawn across the suspended membrane and plot the Fourier transforms of the acquired traces as a function of position in \cref{fig:3}(b). For positions outside the suspended area, phonon damping into the substrate leads to vanishing coherent phonon amplitudes. On the suspended part, we observe three distinct regions (from left to right along the arrow in \cref{fig:3}(a)), with fundamental acoustic breathing mode frequencies of \SI{45}{\GHz}, \SI{43}{\GHz}, and \SI{50}{\GHz}, respectively. Our AFM measurement indicates corresponding slab heights of \SI{25.5}{\nm}, \SI{26.3}{\nm}, and \SI{23.0}{\nm}. We thus deduce a LA phonon speed of sound of \SI{2.3}{\km\per\s} and conclude that a membrane thickness of \SI{60}{\nm} will bring the fundamental, breathing-type phonon resonance to $\sim\SI{20}{\GHz}$, i.e., to a range where we can field-tune the magnon in resonance with the phonon. As in \cref{fig:2}(f), we also observe fainter contributions from the $m=3$ phonon mode which consistently scales with its corresponding fundamental frequency.

\begin{figure}[ht]
	\includegraphics{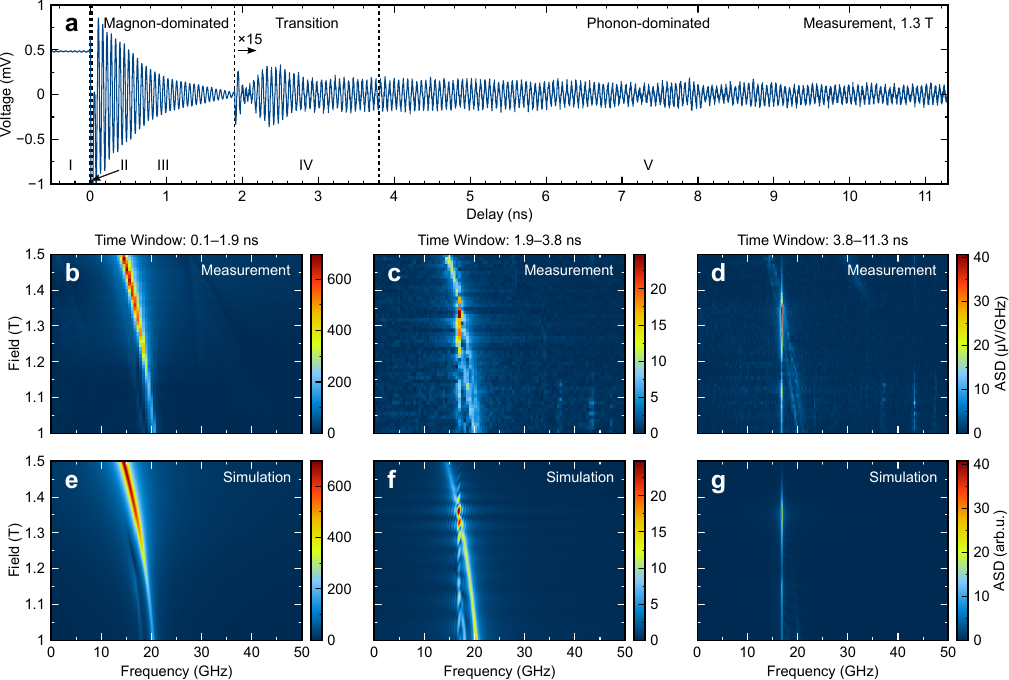}%
	\caption{\label{fig:4}%
		(a)~Measured pump--probe trace with magnon and fundamental breathing-type phonon tuned into resonance by an external magnetic field of \SI{1.3}{\T}. The distinct stages I--V of the dynamics are indicated by dashed lines. Following initial excitation (II), high-momentum magnons quickly escape (III), followed by transitional dynamics (IV) and long-lived phonon oscillations (V). (b--g)~Measured (top row) and simulated (bottom row) spectra of magnetic field dependent pump--probe signals for Fourier time windows of \qtyrange{0.1}{1.9}{\ns} (b, e), \qtyrange{1.9}{3.8}{\ns} (c, f), and \qtyrange{3.8}{11.3}{\ns} (d, g). At short delays signals are dominated by coherent magnons, followed by phonon--magnon interactions at intermediate delays leaving behind the magnon-amplified coherent phonon at long delays for those magnetic field values that bring the two into resonance.}%
\end{figure}

In \cref{fig:4}(a), we show a local pump--probe trace obtained on a sample which we specifically built to fulfill the necessary thickness condition for tuning the out-of-plane magnon and fundamental breathing-type phonon mode in resonance. The specific trace is part of a magnetic field tuning series and was acquired at a field strength of $B=\SI{1.3}{\T}$. The overall dynamics of the measured transient reflectivity $\sfrac{\Delta R(t)}{R_0}$ passes five distinct stages, separated in panel~(a) by dashed vertical lines and labeled I--V:
At $t=0$, the pump pulse impinges on the sample, resonantly excites exciton--polaritons, and leads to the initial steep change in the transient reflectivity $\sfrac{\Delta R(t)}{R_0}$ on time scales corresponding to the effective \SI{400}{\fs} pulse duration (II).
Following the arrival of the pump, coherent oscillations with a period of $\sim\SI{50}{\ps}$ ensue that rapidly decay on a characteristic time scale of \SI{1}{\ns} (III).
Starting at pump--probe delays around \SI{2}{\ns} (IV), we then observe transitional dynamics with clearly developed beatings with a period of \SI{1}{\ns}.
Finally, for pump--probe delays $>\SI{6}{\ns}$ (V), the beating pattern vanishes, leaving a single coherent oscillation behind, whose amplitude decays extremely slowly. This decay is in fact slow enough that a small residual coherent signal from the previous pump pulse remains detectable at negative delays in stage~(I).

To understand the physical mechanisms governing the evolution of $\sfrac{\Delta R(t)}{R_0}$, and in particular identify the distinct signatures of the underlying tripartite quasiparticle interactions, we analyze the pump--probe traces of the magnetic field series in the spectral domain. To this end, we performed separate Fourier transformations for the time windows III--V, exhibiting distinctly different dynamic features. Panels~(b--d) show the resulting amplitude spectral densities (ASD) as a function of magnetic field, with respective delay windows indicated at the top of each panel. As identified in the exemplary time trace of panel~(a), we resolve three distinct stages of the system's dynamics: For early pump--probe delays below \SI{1.9}{\ns} (panel~b), the spectrum is dominated by a strong magnon signature, closely resembling \cref{fig:2}(c). We find two magnon branches: a dominant one at \SI{20}{\GHz} for $B=\SI{1}{\T}$ and a fainter secondary branch shifted to lower frequencies by $\sim\SI{2.5}{\GHz}$. For intermediate delays, between \SI{1.9}{\ns} and \SI{3.8}{\ns} (panel~c), a breathing-type phonon signature additionally appears at a field-independent frequency of \SI{16.8}{\GHz}. At a field of \SI{1.35}{\T}, the magnon and phonon modes are tuned in resonance, and we observe a clear increase in the phonon ASD. Moreover, the phonon ASD in \cref{fig:4}(c) is consistently larger for positive detuning (magnon at higher frequency than phonon) than for negative detuning. At large delays, beyond \SI{3.8}{\ns} (panel~d), the magnon is barely visible, while a clear signature of the phonon mode persists, with its amplitude largest for fields where the magnon was in resonance with the phonon. The same detuning-dependent asymmetry as observed in panel~(c).

We note that the observed coherent magnon dynamics occurs on two very different time scales. The magnon signature in \cref{fig:4}(b) features a linewidth of $\sim\SI{0.5}{\GHz}$, corresponding to an amplitude lifetime of $\tau=\sfrac{\sqrt{3}}{\pi\Delta f_\mathrm{ASD}}=\SI{1}{\ns}$ with $\Delta f_\mathrm{ASD}$ the linewidth obtained from the amplitude spectral density (ASD). This decay time agrees with the decay observed in stage~(III) of panel~(a). In the spectrum for much later delays, corresponding to $4\tau$ to $10\tau$ in panel~(d), however, we still observe a very noticeable magnon contribution, with an ASD only $\approx 40$ times lower than in panel~(b). This factor being much higher than expected from the lifetime extracted in stage~(III) hints at a much longer second decay constant of the magnon. Interestingly, while the magnon signal at large delays is long lived, its linewidth in panel~(d) is nearly equal to the value in panel~(b). To resolve this apparent contradiction, we performed windowed Fourier transforms of the magnetic field series underlying \cref{fig:4} (see \cref{SI-fig:Spectrogram} of the Supporting Information). We find that the long lived magnon component undergoes a delay-dependent frequency blueshift, which broadens its apparent linewidth. 

To understand these observations, in particular the phonon amplification dynamics, we model the system by a set of coupled equations of motion, which we solve to second order in perturbation theory. For the detailed derivation and solution, refer to \cref{SI-sec:Theory} of the Supporting Information. \Cref{fig:4}(e--g) show the simulated magnetic-field dependent ASD in excellent agreement with the experiment. Including only the three interactions outlined in \cref{fig:1}(a), the model accurately reproduces all main features observed in \cref{fig:4}(b--d). The model is based on the uncoupled Hamiltonian
\begin{equation}
	\hat H_0
	=
	\sum_\kk \hbar \omega_{\Pol,\kk} \hat a^\dag_\kk \hat a_\kk
	+
	\sum_\qq \hbar \omega_{\PP,\qq} \hat b^\dag_\qq \hat b_\qq
	+
	\sum_\qq \hbar \omega_{\MM,\qq} \hat c^\dag_\qq \hat c_\qq,
\end{equation}
where $\omega_{\Pol,\kk}$, $\omega_{\PP,\qq}$ and $\omega_{\MM,\qq}$ are the polariton, breathing-type phonon and magnon frequencies, respectively, with the in-plane wave vectors $\kk$ and $\qq$. These wave vectors are good quantum numbers due to the in-plane translational symmetry of the sample. We describe the parametric polariton-–phonon and polariton–-magnon interactions by
\begin{equation}
	\hat H_\mathrm{OM}
	=
	\hbar g_\PP 
	\sum_\kk
	\sum_\qq
	\hat a^\dag_\kk \hat a_{\kk+\qq} 
	(\hat b_\qq^\dag + \hat b_{-\qq})
	+
	\hbar g_\MM
	\sum_\kk
	\sum_\qq
	\hat a^\dag_\kk \hat a_{\kk+\qq} 
	(\hat c_\qq^\dag + \hat c_{-\qq}),
\end{equation}
with the parametric coupling constants $g_\PP$ and $g_\MM$ as introduced in \cref{fig:1}(a). The Hamiltonian also includes the resonant coupling between magnons and phonons arising from magnetoelastic coupling via a beam splitter-type interaction of strength $V$: 
\begin{equation}
	\hat H_\mathrm{PM}
	=
	\hbar V \sum_\qq(\hat c_\qq^\dag \hat b_\qq + \hat b^\dag_\qq \hat c_\qq)
	.
\end{equation}
The sign of the product $V g_\PP g_\MM$ determines whether the direct and magnon-mediated excitation pathways of the phonon mode interfere constructively or destructively for a given sign of detuning (see \cref{SI-analytic spectrum,SI-DFT,SI-reflectivity} of the Supporting Information). The pronounced asymmetry of the phonon amplitude with respect to
magnon--phonon detuning in the experimental data thus directly indicates an attractive
interaction, $V<0$ (for $g_\PP, g_\MM>0$), which leads to phonon
amplification for positive detuning, in agreement with the model in
\cref{fig:4}(f,g). Finally, the interaction of the system with the external pump and probe fields through the polariton states is introduced as Gaussian pulses in space and time, which drive a coherent polariton amplitude. We then solve the resulting set of coupled equations of motion for the expectation values of the polariton, magnon, and phonon amplitudes, with all three set to zero as the initial condition. Details are compiled in \cref{SI-sec:Theory} of the Supporting Information. In these equations of motion, we use realistic, momentum-dependent dissipation rates for all three interacting quasiparticles. The main contribution to the dissipation of the coherent magnons $\gamma_{\MM,\qq}$ stems from kinetic losses, namely their leakage out of the probe region~\cite{sun2024a}. We therefore estimate $\gamma_{\MM,\qq} = L v_{\mathrm{g},\qq}/S$, where $L$ is the circumference of the probe region, $S$ the probed area, and $v_{\mathrm{g},\qq} = \abs{\pdv*{\omega_{\MM,\qq}}{\qq}}$ the magnon group velocity. Assuming that the probe region is a circular spot with diameter $d$, we obtain $\gamma_{\MM,\qq}=\frac{4}{d}\abs{\pdv*{\omega_{\MM,\qq}}{\qq}}$. We consider the parametric coupling constants $g_\alpha$ as small parameters and find the solution to the equations of motion in the form of a perturbation series in these coupling constants~\cite{lisjanskij2024}. 

Up to zeroth order, the equations of motion describe the resonant excitation of polaritons by the pump and probe pulses. The physical meaning of this solution is an elastic scattering, equivalent to the Rayleigh contribution in the quantum theory of light scattering. In first order, the equations of motion capture the parametric excitation of coherent magnons and coherent phonons by the excited polaritons. In second order, the equations of motion capture the influence of the parametrically excited magnons and phonons on the coherently oscillating polaritons, which are back-action effects seen by the probe pulse. Assuming the direct magnon--phonon coupling to be small enough such that $V < \abs{\gamma_{\MM,\qq} - \gamma_{\PP,\qq}}$, the direct resonant coupling also becomes a second-order effect. From the second order perturbation theory solution, we extract the impact of the coherent dynamics of phonons and magnons on the probe pulse, revealing the modulation in its amplitude. If the change in reflectivity $\Delta R$ is small, we can approximate $\Delta R/R_0 \approx 2 \Re\left[E_0^{*} \Delta E\right]/\abs{E_0}^2$. The fields $E_0$ and $\Delta E$ are determined by the response of the sample, therefore $E_0 \propto a^{(0)}_\kk$ for zero pump pulse amplitude and $\Delta E \propto a^{(2)}_\kk$ in the presence of the pump. With the only further assumptions that the optical pulses are short on the time scale of the coherent magnon and phonon oscillations, and a weak polariton dispersion over the relevant momentum range, we find a closed formula (Supporting Information \cref{SI-analytic spectrum}) for the ASD evaluated over a rectangular Fourier window. This is the equation underlying the plots in \cref{fig:4}(e--g). To match the result of the perturbation theory to the experiment, we first fitted the magnetic field dependent magnon dispersion and linewidth as well as the interaction constant $g_\MM$ from \cref{fig:4}(b). We then extracted the breathing-type phonon frequency and linewidth from panel~(d) (see Supporting Information \cref{SI-fig:Parameters}).

\begin{figure}[ht]
	\includegraphics{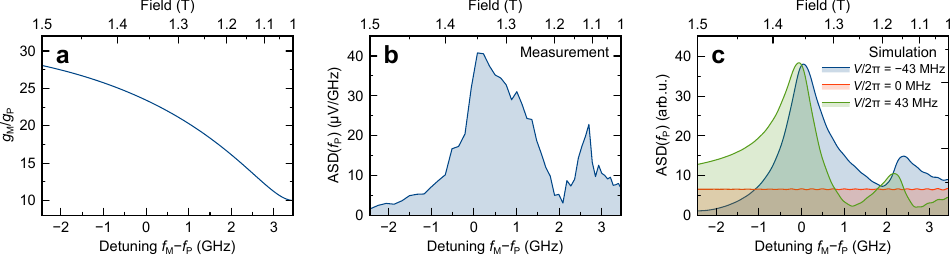}%
	\caption{\label{fig:5}%
		(a)~Simulated parametric coupling ratio $\sfrac{g_\MM}{g_\PP}$ as a function of detuning between the phonon frequency $f_\PP=\SI{16.8}{\GHz}$ and the dominant magnon branch frequency $f_\MM$. (b)~Measured spectra cross section of \cref{fig:4}(d) at $f = f_\PP$ plotted versus magnon--phonon detuning. (c)~Simulated spectra cross section of \cref{fig:4}(g) at $f = f_\PP$. In addition to the case of attractive magnon--phonon coupling strength $V/2\pi = \SI{-43}{\MHz}$ (blue) which best matches the experimental result in (b), the cases $V/2\pi = \SI{0}{\MHz}$ (orange) and $V/2\pi = \SI{43}{\MHz}$ (green) are shown for reference.}%
\end{figure}

As an important first result inferred from the simulations, we obtain the ratio of parametric magnon-to-polariton and phonon-to-polariton coupling constants $\sfrac{g_\MM}{g_\PP}$ as plotted in \cref{fig:5}(a) as a function of the detuning between the coherent phonon frequency $f_\PP=\SI{16.8}{\GHz}$ and the magnon frequency $f_\MM$ of the dominant branch. While the ratio varies by a factor of two over the explored detuning range, the coherent magnons are generally excited with displacements one order of magnitude larger than the coherent phonons. 

Importantly, we extracted the interaction strength $V$ between phonons and magnons by considering the FFT spectra in \cref{fig:4}(d) over the full range of the magnetic field. This is the only independently fitted parameter in our model, and we obtain the value $V/2\pi = \SI{-43}{\MHz}$. 
The negative sign of the magnon--phonon interaction follows directly from the detuning dependence of the phonon signal. In \cref{fig:5}(b), we plot the cross-section of \cref{fig:4}(d) at $f_\PP$ as a function of magnon--phonon detuning. The phonon amplitude is resonantly enhanced near zero detuning, approaching \SI{40}{\uV\per\GHz}, and the lineshape is strongly asymmetric: at $+\SI{1}{\GHz}$ (magnon above the phonon), the signal remains at about \SI{30}{\uV\per\GHz}, whereas at $-\SI{1}{\GHz}$ it rapidly drops down below \SI{6}{\uV\per\GHz}. The simulation in \cref{fig:5}(c) shows that this asymmetry is a direct measure of the sign of $V$. Without the interaction ($V=0$, orange), the signal at $f_\PP$ is independent of the detuning, so the resonance itself originates from magnon--phonon coupling. For $g_\PP>0$ and $g_\MM>0$, a repulsive interaction ($V/2\pi = \SI{43}{\MHz}$, green) enhances the phonon on the negative-detuning side and suppresses it on the positive side, opposite to the observation. Only the attractive magnon--phonon interaction ($V/2\pi = \SI{-43}{\MHz}$, blue) is able to reproduce the experimental signatures, including the enhancement at zero detuning, the asymmetry towards positive detuning, and the dip and secondary maximum at \qtyrange{2}{3}{\GHz}.

Based on the excellent fit between the experiment and our model, the physical interpretation of the pump--probe measurements is the following: The pump pulse excites both magnons and phonons. The magnons are, however, excited much more efficiently. In the time window \qtyrange{0.1}{1.9}{\ns}, most of the magnons with high in-plane momenta escape the probe region as their group velocity is as high as several kilometers per second~\cite{sun2024a}. This results in the fast decay of $\Delta R(t)/R_0$ at short delays. The coherent magnons transfer part of their energy to the coherent phonons with a frequency of \SI{16.8}{\GHz}. The phonons excited due to this process are not only localized, but also have a very low dissipation rate compared to the escape rate of the coherent magnons and are hence still present at very long delays.
Alongside the phonons, we also observe long-lived magnons in \cref{fig:4}(d). These low-momentum magnons have a low group velocity and are thus unable to escape the probe region~\cite{sun2024a}. Due to their relatively small amplitude, the resonant magnon--phonon interaction is, however, dominated by the high-momentum components present at early delays.

\section{Conclusion}

In summary, we have demonstrated optically detected tripartite polariton--magnon--phonon dynamics in suspended CrSBr membranes with engineered, high-$Q$ breathing modes. Combining resonant polariton pump--probe spectroscopy with magnetic-field tuning of the magnon resonance allows us to quantify the relative polariton--magnon and polariton--phonon interaction strengths, $\sfrac{g_\MM}{g_\PP}$, directly from the measured dynamics. This provides a practical route to determining $g_\MM$ once $g_\PP$ is obtained from first-principles calculations. Most importantly, we determine a magnon--phonon interaction strength of $V/2\pi = \SI{-43}{\MHz}$.
Its negative sign, corresponding to an attractive interaction, follows directly from the asymmetric detuning dependence of the phonon amplification, which arises from constructive interference between the direct and the magnon-mediated phonon excitation. At resonance, energy flows almost unidirectionally from the rapidly escaping magnons into the long-lived breathing mode.
Our results establish suspended CrSBr as a platform for interfacing coherent spin, lattice, and optical excitations with engineered gigahertz acoustic resonances and optical readout.
Because the phonon response is governed by interference between two excitation pathways, its magnitude is set by the relative phase of the magnon and phonon excitation, which could be controlled coherently. Followup experiments may include pump-pump-probe configurations, where the delay between two pump pulses will enable coherently switching the magnon-to-phonon energy transfer on and off.
Moreover, $\abs{V}$ is not too far from the hybridization condition $\abs{\gamma_\MM-\gamma_\PP}/2$ and the magnon--phonon back-action is limited only by the kinetic losses of the magnons. Reducing these losses by lateral patterning of the CrSBr membrane should therefore drive the system through an exceptional point, from unidirectional energy transfer into the strong-coupling regime of coherent, bidirectional magnon--phonon exchange, a prerequisite for quantum-coherent magnon--phonon transduction in our platform.

\begin{acknowledgments}
	This project was funded by the \begin{otherlanguage}{ngerman}Deutsche Forschungsgemeinschaft\end{otherlanguage} {} (DFG, German Research Foundation), grant numbers INST 184/249-1 FUGG and funded in part by the Deutsche Forschungsgemeinschaft (DFG, German Research Foundation) under Germany’s Excellence Strategy - EXC 3051/1 ``NaviSense'' project number 533653176.
	The authors acknowledge support by the European Commission (ERC Dual-Twist, Grant number 101170213) and through the \begin{otherlanguage}{ngerman}Niedersächsisches Ministerium für Wissenschaft und Kultur\end{otherlanguage} {} (Wissenschaftsraum ElLiKo within the programme ``zukunft.niedersachsen''). We acknowledge support by Jane and Aatos Erkko Foundation, by the Finnish Research Impact Foundation within the Tandem Industry Academia (TIA) Seed Project no. 667,  and by Photonics Research and Innovation Flagship (PREIN) through the SEED Collaborative Projects and Incentive Scheme at Aalto University. Sh.V.Yu. thanks the Magnus Ehrnrooth foundation. 
	Z.S.\ was supported by project LUAUS25268 from Ministry of Education Youth and Sports (MEYS).
\end{acknowledgments}

\bibliography{local}

\end{document}

\title{\texorpdfstring{Supporting Information\\Magnon--Phonon--Polariton Optomechanics in a van der Waals Magnet}{Supporting Information -- Magnon--Phonon--Polariton Optomechanics in a van der Waals Magnet}}

\author{Jens-Christian Drawer}%
\affiliation{Institut für Physik, Fakultät V, Carl von Ossietzky Universität Oldenburg, 26129 Oldenburg, Germany}%

\author{Vladislav Yu.\ Shishkov}%
\affiliation{Macroscopic Quantum Optics (MQO) Labs, Department of Applied Physics, Aalto University School of Science, FI-00076 Espoo, Finland}%

\author{Bo Han}%
\affiliation{Institut für Physik, Fakultät V, Carl von Ossietzky Universität Oldenburg, 26129 Oldenburg, Germany}%

\author{Vita Solovyeva}%
\affiliation{Institut für Physik, Fakultät V, Carl von Ossietzky Universität Oldenburg, 26129 Oldenburg, Germany}%

\author{Iva Plutnarová}%
\affiliation{Department of Inorganic Chemistry, Faculty of Chemical Technology, University of Chemistry and Technology Prague, Technick\'{a} 5, Prague 6, 16628, Czech Republic}%

\author{Zdeněk Sofer}%
\affiliation{Department of Inorganic Chemistry, Faculty of Chemical Technology, University of Chemistry and Technology Prague, Technick\'{a} 5, Prague 6, 16628, Czech Republic}%

\author{Christian Schneider}%
\affiliation{Institut für Physik, Fakultät V, Carl von Ossietzky Universität Oldenburg, 26129 Oldenburg, Germany}%

\author{Anton V.\ Zasedatelev}%
\affiliation{Macroscopic Quantum Optics (MQO) Labs, Department of Applied Physics, Aalto University School of Science, FI-00076 Espoo, Finland}%

\author{Martin Esmann}%
\thanks{Corresponding author: \href{mailto:m.esmann@uni-oldenburg.de}{m.esmann@uni-oldenburg.de}}
\affiliation{Institut für Physik, Fakultät V, Carl von Ossietzky Universität Oldenburg, 26129 Oldenburg, Germany}%

\date{\today}%

\maketitle

\renewcommand{\thefigure}{S\arabic{figure}}
\renewcommand{\theequation}{S\arabic{equation}}

\FloatBarrier
\section{\label{sec:ExperimentalMethods}Experimental methods -- Pump--probe spectroscopy}

\begin{figure}[ht]
	\includegraphics{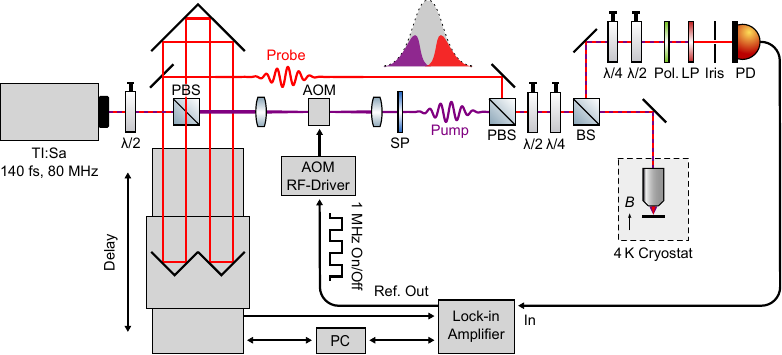}%
	\caption{\label{fig:Setup}%
		Schematic of the pump--probe setup.}%
\end{figure}

A sketch of the experimental pump--probe setup is shown in \cref{fig:Setup}. The sample is held at \SI{4}{\K} in a closed-cycle cryostat (AttoCube AttoDry1000) on an XYZ translation stage. Optical imaging, as well as cross-polarized optical pump--probe spectroscopy, are all performed through the same cold microscope objective (Thorlabs 354105-B with $\mathrm{NA}=\num{0.6}$). We derive optical pulses with \SI{140}{\fs} nominal duration at a repetition rate of \SI{80}{\MHz} and \SI{905}{\nm} central wavelength from a Ti:Sapphire laser (Coherent Chameleon Ultra II). By passing the laser through a polarizing beam splitter (PBS), we prepare two orthogonally polarized pulse trains. The relative power of the two pulse trains is controlled with a $\sfrac{\lambda}{2}$-waveplate in front of the PBS. We send the train of probe pulses through a motorized \SI{60}{\cm} long delay line (Newport M-IMS600LM). By folding the beam into six passes using hollow retro-reflectors (Newport), we obtain a maximum optical delay of \SI{12}{\ns}, i.e., we can explore the whole dynamics between subsequent pump events. The train of pump pulses is passed through an acousto-optical modulator (AOM, Gooch \& Housego 3080-125), imprinting a square-wave modulation on the pump at a frequency of \SI{1}{\MHz} and \SI{50}{\percent} duty cycle. To facilitate spectral separation of pump and probe beam in the collection, the pump is filtered at its center with a shortpass filter (Semrock \SI{995}{\nm} VersaChrome Edge™ tunable shortpass filter). The two cross polarized beams are recombined in a second PBS, adjusted in their polarization with a set of $\sfrac{\lambda}{2}$ and $\sfrac{\lambda}{4}$-waveplates to align with the sample crystal axes, and focused tightly onto the sample to an estimated focal diameter of $\sim$\SI{1}{\um}. Typical optical powers impinging on the sample were \SI{1}{\mW} for the pump and \SI{0.1}{\mW} for the probe beam. The reflected light is passed through a second set of $\sfrac{\lambda}{2}$ and $\sfrac{\lambda}{4}$-waveplates in combination with a polarizer, followed by longpass filter (Semrock \SI{995}{\nm} VersaChrome Edge™ tunable longpass filter) and an iris, allowing us to suppress the pump pulses. Finally, the reflected probe impinges onto a fast photo diode (Femto OE-300-IN-01). We demodulate the photo diode signal on the fundamental AOM frequency using a lock-in amplifier (Zurich Instruments MFLI). The probe's delay-dependent differential reflectivity time trace is extracted. Its incoherent background is then removed by fitting and subtracting a spline of limited roughness, which was chosen low enough to result in minimal impact on the relevant part of the spectrum. Depending on the following processing of the data, any remaining incoherent background can be further removed by fitting and subtracting a spline of limited roughness, which was chosen low enough to result in minimal impact on the spectrum at and beyond the fundamental mode. Finally, spectra are obtained by calculating the Fourier transform of the background-free time trace. If required to improve visibility of high-frequency components, we may take the first derivative with respect to time, i.e., we calculate the difference between subsequent equidistant voltage values before performing the Fourier transform. This multiplies the amplitude spectrum by a factor of $\omega$.

\FloatBarrier
\section{\label{sec:Theory}Theoretical modeling}

The relative reflectivity change $\Delta R(t)/R_0$ is related to the electromagnetic field $E$ reflected from the sample as
\begin{equation}
	\frac{\Delta R(t)}{R_0} 
	=
	\frac{\abs{E_0(t) + \Delta E(t)}^2 - \abs{E_0(t)}^2}{\abs{E_0(t)}^2},
\end{equation}
where $E_0(t)$ is the electric field of the reflected probe pulse in the absence of the pump and $\Delta E(t)$ the additional component when the pump is switched on.
When $\abs{\Delta E(t)} \ll \abs{E_0(t)}$, we obtain
\begin{equation}
	\frac{\Delta R(t)}{R_0} 
	=
	2\frac{\Re\left[E_0^*\Delta E(t)\right]}{\abs{E_0}^2}.
\end{equation}

The Hamiltonian of polaritons, magnons, and phonons is
\begin{equation}
	\hat H_0
	=
	\sum_\kk \hbar \omega_{\Pol,\kk} \hat a^\dag_\kk \hat a_\kk
	+
	\sum_\qq \hbar \omega_{\PP,\qq} \hat b^\dag_\qq \hat b_\qq
	+
	\sum_\qq \hbar \omega_{\MM,\qq} \hat c^\dag_\qq \hat c_\qq,
\end{equation}
where $\omega_{\Pol,\kk}$ are the natural frequencies of the polaritons with in-plane wave vector $\kk$, $\omega_{\PP,\qq}$ and $\omega_{\MM,\qq}$ are the breathing-type phonon mode and magnon frequencies, respectively, with the in-plane wave vector $\qq$. Note that the aforementioned wave vectors are good quantum numbers due to the in-plane translational symmetry of the sample.

The Hamiltonian of the optomechanical interaction between polaritons and magnons, and between polaritons and phonons, is
\begin{equation}
	\hat H_\mathrm{OM}
	=
	\hbar g_\PP 
	\sum_\kk
	\sum_\qq
	\hat a^\dag_\kk \hat a_{\kk+\qq} 
	(\hat b_\qq^\dag + \hat b_{-\qq})
	+
	\hbar g_\MM
	\sum_\kk
	\sum_\qq
	\hat a^\dag_\kk \hat a_{\kk+\qq} 
	(\hat c_\qq^\dag + \hat c_{-\qq}),
\end{equation}
where $g_\PP$ is the parametric coupling constant between phonons and polaritons, and $g_\MM$ is the parametric coupling constant between magnons and polaritons.

The interaction term in the Hamiltonian describing the resonant coupling between magnons and phonons is expressed as 
\begin{equation}
	\hat H_\mathrm{PM}
	=
	\hbar V \sum_\qq(\hat c_\qq^\dag \hat b_\qq + \hat b^\dag_\qq \hat c_\qq)
	,
\end{equation}
where $V$ is the interaction strength between phonons and magnons.

The interaction with the external fields (pump and probe drives) through polariton states is introduced by
\begin{equation}
	\hat H_\textrm{Pump--Probe}
	=
	\sum_\kk
	\hbar W_{\Pump,\kk}(t)
	\left(
	\hat a_\kk \ee^{\ii\omega_\Pump t} + \hat a^\dag_\kk \ee^{-\ii\omega_\Pump t}
	\right)
	+
	\hbar W_{\Probe,\kk}(t)
	\left(
	\hat a_\kk \ee^{\ii\omega_\Probe t} + \hat a^\dag_\kk \ee^{-\ii\omega_\Probe t}
	\right),
\end{equation}
where
\begin{align}
	W_{\Pump,\kk}(t)
	&=
	\Omega_{\Pump}
	\ee^{-t^2/2\tau_\Pump^2}
	\frac{1}{2\pi\sigma^2}
	\ee^{-\kk^2/2\sigma^2},
	\\
	W_{\Probe,\kk}(t)
	&=
	\Omega_{\Probe}
	\ee^{-(t-\Delta t)^2/2\tau_\Probe^2}
	\frac{1}{2\pi\sigma^2}
	\ee^{-\kk^2/2\sigma^2},
\end{align}
and we assume that both pump and probe are Gaussian beams with a width in momentum space of $\sigma$.

\subsection{Analytical solution}

The equations of motion for the polariton, magnon, and phonon amplitudes $A = \langle \hat A \rangle$ are
\begin{multline}
	\dv{a_\kk}{t}
	=
	\left(
	-\ii \omega_{\Pol,\kk} - \frac{\gamma_{\Pol,\kk}}{2}
	\right)
	a_\kk
	-\ii g_\PP \sum_\qq a_{\kk+\qq} (b^*_\qq + b_{-\qq})
	-\ii g_\MM \sum_\qq a_{\kk+\qq} (c^*_\qq + c_{-\qq})
	\\
	-\ii W_{\Pump,\kk}(t) \ee^{-\ii\omega_\Pump t} 
	-\ii W_{\Probe,\kk}(t) \ee^{-\ii\omega_\Probe t},
\end{multline}
\begin{align}
	\dv{b_\qq}{t}
	&=
	\left(
	-\ii \omega_{\PP,\qq} - \frac{\gamma_{\PP,\qq}}{2}
	\right)
	b_\qq
	-\ii V c_\qq
	-ig_\PP 
	\sum_\kk
	\hat a^\dag_\kk \hat a_{\kk+\qq},
	\\
	\dv{c_\qq}{t}
	&=
	\left(
	-\ii \omega_{\MM,\qq} - \frac{\gamma_{\MM,\qq}}{2}
	\right)
	c_\qq
	-\ii V b_\qq
	-\ii g_\MM 
	\sum_\kk
	\hat a^\dag_\kk \hat a_{\kk+\qq}.
\end{align}
Here, we also introduced $\gamma_{\Pol,\kk}$ as the dissipation rate of the polaritons, $\gamma_{\PP,\qq}$ as the dissipation rate of the phonons, and $\gamma_{\MM,\kk}$ as the dissipation rate of the magnons.

The main contribution to the dissipation of the coherent magnons stems from kinetic losses, namely their leakage out of the probe region~\cite{sun2024a}. Thus, we can express the magnon dissipation as $\gamma_{\MM,\qq} = L v_{\mathrm{g},\qq}/S$, where $L$ is the circumference of the probe region, $S$ the probed area, and $v_{\mathrm{g},\qq} = \abs{\pdv*{\omega_{\MM,\qq}}{\qq}}$ the magnon group velocity.
Assuming that the probe region is a circular spot with diameter $d$, we obtain
\begin{equation}
	\gamma_{\MM,\qq} 
	=
	\frac{4}{d}
	\abs{\pdv{\omega_{\MM,\qq}}{\qq}}.
\end{equation}

We set the initial conditions
\begin{align}
	a_\kk{(0)} &= 0,\\
	b_\qq{(0)} &= 0,\\
	c_\qq{(0)} &= 0,
\end{align}

i.e., zero polariton, magnon and phonon amplitude before the arrival of the pump pulse. We consider the parametric coupling constants $g_\alpha$ as small parameters for the perturbation theory~\cite{lisjanskij2024}.
We find the solution to the equations of motion for the amplitudes in the form of series in this coupling constants $A = A^{(0)} + g_\alpha A^{(1)} + g^2_\alpha A^{(2)} + \ldots$.

\subsubsection{0th order of the perturbation theory}

The equations for the zeroth order of the perturbation theory are
\begin{align}
	\dv{a^{(0)}_\kk}{t}
	&=
	\left(
	-\ii \omega_{\Pol,\kk} - \frac{\gamma_{\Pol,\kk}}{2}
	\right)
	a^{(0)}_\kk
	-\ii W_{\Pump,\kk}(t) \ee^{-\ii\omega_\Pump t}
	-\ii W_{\Probe,\kk}(t) \ee^{-\ii\omega_\Probe t},
	\\
	\dv{b^{(0)}_\qq}{t}
	&=
	\left(
	-\ii \omega_{\PP,\qq} - \frac{\gamma_\PP}{2}
	\right)
	b^{(0)}_\qq
	-\ii V c^{(0)}_\qq,
	\\
	\dv{c^{(0)}_\qq}{t}
	&=
	\left(
	-\ii \omega_{\MM,\qq} - \frac{\gamma_{\MM,\qq}}{2}
	\right)
	c^{(0)}_\qq
	-\ii V b^{(0)}_\qq.
\end{align}

The solution of these equations is
\begin{align}
	a^{(0)}_\kk 
	&\approx 
	A_{\Pump,\kk}(t) \ee^{-\ii\omega_\Pump t}
	+
	A_{\Probe,\kk}(t) \ee^{-\ii\omega_\Probe t},
	\\
	b^{(0)}_\qq &= 0,
	\\
	c^{(0)}_\qq &= 0,
\end{align}
where we denoted the momentum-resolved, time-dependent polariton amplitudes
\begin{align}
	A_{\Pump,\kk}(t) 
	&= 
	\frac{ 
		W_{\Pump,\kk}(t)  
	}{
		\omega_{\Pump} - \omega_{\Pol,\kk} + \ii \gamma_{\Pol,\kk}/2
	}
	,
	\\
	A_{\Probe,\kk}(t) 
	&= 
	\frac{ 
		W_{\Probe,\kk}(t)  
	}{
		\omega_\Probe - \omega_{\Pol,\kk} + \ii \gamma_{\Pol,\kk}/2
	}
	.
\end{align}

The physical meaning of this solution is an elastic scattering, equivalent to the Rayleigh contribution in the quantum theory of light scattering. 

\subsubsection{1st order of the perturbation theory}

Assuming that the pump intensity is much higher than the probe intensity, and taking into account the solutions for zeroth order of the perturbation theory, we obtain the equations for the first order of the perturbation theory
\begin{align}
	\dv{a^{(1)}_\kk}{t}
	&=
	\left(
	-\ii \omega_{\Pol,\kk} - \frac{\gamma_{\Pol,\kk}}{2}
	\right) 
	a_\kk^{(1)},
	\\
	\dv{b^{(1)}_\qq}{t}
	&=
	\left(
	-\ii \omega_{\PP,\qq} - \frac{\gamma_{\PP,\qq}}{2}
	\right)
	b^{(1)}_\qq
	-\ii V c^{(1)}_\qq
	-\ii g_\PP \sum_\kk 
	a^{(0)*}_\kk a^{(0)}_{\kk+\qq},
	\\
	\dv{c^{(1)}_\qq}{t}
	&=
	\left(
	-\ii \omega_{\MM,\qq} - \frac{\gamma_{\MM,\qq}}{2}
	\right)
	c^{(1)}_\qq
	-\ii V b^{(1)}_\qq
	-\ii g_\MM 
	\sum_\kk
	a^{(0)*}_\kk a^{(0)}_{\kk+\qq}.
\end{align}

We assume that the pump duration $\tau_\Pump$ is much shorter than the period of the magnons and phonons, that is, $\tau_\Pump \omega_\MM \ll 1$ and $\tau_\Pump \omega_\PP \ll 1$, so we can replace

\begin{equation} \label{pump envelope approximation}
	\ee^{-t^2/2\tau_{\Pump}^2} 
	\approx 
	\sqrt{2\pi} \tau_{\Pump} \delta(t).
\end{equation}

We also use
\begin{equation}
	\sum_\kk 
	\frac{1}{2\pi\sigma^2}
	\ee^{-\kk^2/2\sigma^2}
	\frac{1}{2\pi\sigma^2}
	\ee^{-(\kk+\qq)^2/2\sigma^2}
	=
	\frac{S}{4\pi^2}\frac{\ee^{-\qq^2/4\sigma^2}}{4\pi\sigma^2}
	=
	\frac{Se^{-\qq^2/4\sigma^2}}{16\pi^3\sigma^2}.
\end{equation}
Thus in the equations above
\begin{equation}
	\sum_\kk a^{(0)*}_\kk a^{(0)}_{\kk+\qq}
	\approx
	\frac
	{
		\sqrt{2\pi}\tau_\Pump
		\Omega_{\Pump}^2
	}{
		(\omega_\Pump - \omega_{\Pol,\kk})^2 + \gamma_{\Pol,\kk}^2/4
	}
	\frac{Se^{-\qq^2/4\sigma^2}}{16\pi^3\sigma^2}
	\delta(t).
\end{equation}

The solution of the equations is
\begin{align}
	a_\kk^{(1)} &= 0,
	\\
	b^{(1)}_\qq
	&= 
	( B_\qq - v_\qq C_\qq ) \ee^{-(\ii\omega_{\PP,\qq}+\gamma_{\PP,\qq}/2)t} +  
	v_\qq C_\qq \ee^{-(\ii\omega_{\MM,\qq}+\gamma_{\MM,\qq}/2)t},
	\\
	c^{(1)}_\qq  
	&= 
	( C_\qq + v_\qq B_\qq ) \ee^{-(\ii\omega_{\MM,\qq}+\gamma_{\MM,\qq}/2)t} -  
	v_\qq B_\qq \ee^{-(\ii\omega_{\PP,\qq}+\gamma_{\PP,\qq}/2)t},
\end{align}
where we assumed a weak coupling between magnons and phonons, such that $V < \abs{\gamma_{\MM,\qq} - \gamma_{\PP,\qq}}$, and denoted
\begin{align}
	B_\qq
	&=
	-
	\ii
	\frac
	{
		g_\PP
		\sqrt{2\pi} \tau_\Pump
		\Omega_\Pump^2
	}{
		(\omega_\Pump - \omega_{\Pol,\qq})^2 + \gamma_{\Pol,\qq}^2/4
	}
	\frac{Se^{-\qq^2/4\sigma^2}}{16\pi^3\sigma^2}
	,
	\\
	C_\qq
	&=
	-\ii
	\frac
	{
		g_\MM
		\sqrt{2\pi} \tau_\Pump
		\Omega_{\Pump}^2
	}{
		(\omega_\Pump - \omega_{\Pol,\qq})^2 + \gamma_{\Pol,\qq}^2/4
	}
	\frac{Se^{-\qq^2/4\sigma^2}}{16\pi^3\sigma^2}
	,
	\\
	v_\qq
	&=
	\frac
	{ \ii V  
	}{
		\ii ( \omega_{\MM,\qq} - \omega_{\PP,\qq} ) + (\gamma_{\MM,\qq} - \gamma_{\PP,\qq})/2
	}.
\end{align}

The physical meaning of the solution for the first order of the perturbation theory is a parametric excitation of coherent magnons and phonons.

\subsubsection{2nd order of the perturbation theory}

In the second order of the perturbation theory, we consider the influence of the parametrically excited magnons and phonons on the coherently oscillating polaritons, which are back-action effects.
Here, it is sufficient to consider only the equation for $a_\kk^{(2)}$, given by
\begin{equation}
	\dv{a_\kk^{(2)}}{t}
	=
	\left(
	-\ii \omega_{\Pol,\kk} - \frac{\gamma_{\Pol,\kk}}{2}
	\right)
	a_\kk^{(2)}
	-\ii g_\PP \sum_\qq a_{\kk+\qq}^{(0)} (b^{*(1)}_\qq + b^{(1)}_{-\qq})
	-\ii g_\MM \sum_\qq a_{\kk+\qq}^{(0)} (c^{*(1)}_\qq + c^{(1)}_{-\qq}),
\end{equation}
where we took into account all the solutions for all the previous orders of the perturbation theory.

We focus on the time after the arrival of the probe pulse, $t > \Delta t$, analyzing the change in the scattered probe field caused by parametrically excited coherent oscillations of magnons and phonons.
In this case, the equation for $a_\kk(t)$ in the second order of the perturbation theory becomes
\begin{equation}
	\dv{a_\kk^{(2)}}{t}
	=
	\left(
	-\ii \omega_{\Pol,\kk} - \frac{\gamma_{\Pol,\kk}}{2}
	\right)
	a_\kk^{(2)}
	-\ii g_\PP \sum_\qq A_{\Probe,\kk+\qq}(t) (b^{*(1)}_\qq + b^{(1)}_{-\qq}) \ee^{-\ii\omega_\Probe t}
	-\ii g_\MM \sum_\qq A_{\Probe,\kk+\qq}(t) (c^{*(1)}_\qq + c^{(1)}_{-\qq}) \ee^{-\ii\omega_\Probe t}
\end{equation}

Assuming that $\omega_{\MM,\kk} < \gamma_{\Pol,\kk}$ and $\omega_{\PP,\kk} < \gamma_{\Pol,\kk}$, we write the approximate solution to this equation
\begin{equation}
	a_\kk^{(2)}
	\approx
	\Delta A_{\Probe,\kk}(t)
	\ee^{-\ii \omega_\Probe t},
\end{equation}
where
\begin{multline}
	\Delta A_{\Probe,\kk}(t)
	\approx
	\sum_\qq
	\frac{ A_{\Probe,\kk+\qq}(t) }{ \omega_\Probe - \omega_{\Pol,\kk} + \ii \gamma_{\Pol,\kk}/2 }
	\\
	\left\{
	\left[
	B_\qq (g_\PP - v_\qq g_\MM) - C_\qq v_\qq g_\PP
	\right]
	\ee^{-(\ii\omega_{\PP,\qq} + \gamma_{\PP,\qq}/2 )t}
	+
	\left[
	C_\qq (g_\MM + v_\qq g_\PP) + B_\qq v_\qq g_\MM
	\right]
	\ee^{-(\ii\omega_{\MM,\qq} + \gamma_{\MM,\qq}/2 )t}
	+
	\mathrm{h.c.}
	\right\}
	.
\end{multline}

From the second order perturbation theory, we extract an impact of excited coherent dynamics of phonons and magnons on the probe pulse, revealing the modulation in its amplitude.

As discussed above, $\Delta R/R_0 \approx 2 \Re\left[E_0^{*} \Delta E\right]/\abs{E_0}^2$.
The fields $E_0$ and $\Delta E$ are determined by the response of the sample, therefore $E_0 \propto a^{(0)}_\kk$ at $\Omega_\Pump=0$ and $\Delta E \propto a^{(2)}_\kk$ in the presence of the pump field.
Given that we measure intensity integrated over the surface of the detector, we obtain
\begin{equation}
	E_0^* \Delta E 
	\propto 
	\sum_\kk a^{(0)*}_\kk a^{(2)}_\kk
	=
	\sum_\kk A_{\Probe,\kk}^*(t) \Delta A_{\Probe,\kk}(t).
\end{equation}

Assuming that $\omega_{\Pol,\kk}\approx\omega_\Pol=\mathrm{const}$ and $\gamma_{\Pol,\kk}\approx\gamma_\Pol=\mathrm{const}$ for the wave vector excited by the pump pulse, $\abs{\kk} < \sigma$, we obtain 
\begin{multline}
	\sum_\kk A_{\Probe,\kk}^*(t) \Delta A_{\Probe,\kk}(t)
	=
	\frac
	{ 
		\Omega_{\Probe}^2 \ee^{-(t-\Delta t)^2/\tau_\Probe^2}
	}
	{ 
		(\omega_\Probe - \omega_{\Pol})^2 + \gamma_{\Pol}^2/4 
	}
	\frac{S}{16\pi^3\sigma^2}
	\\
	\sum_\qq
	\ee^{-\qq^2/4\sigma^2}
	\left\{
	\left[
	B_\qq (g_\PP - v_\qq g_\MM) - C_\qq v_\qq g_\PP
	\right]
	\ee^{-(\ii\omega_{\PP,\qq} + \gamma_{\PP,\qq}/2 )t}
	+
	\left[
	C_\qq (g_\MM + v_\qq g_\PP) + B_\qq v_\qq g_\MM
	\right]
	\ee^{-(\ii\omega_{\MM,\qq} + \gamma_{\MM,\qq}/2 )t}
	+
	\mathrm{h.c.}
	\right\}
	.
\end{multline}

Using the replacement~(\ref{pump envelope approximation}), but this time for the probe pulse, we obtain
\begin{equation}
	\frac{\Delta R(t)}{R_0}
	=
	I_0 \Im
	\left\{
	\frac{2\pi}{S\sigma^2}\sum_\qq
	\ee^{-\qq^2/2\sigma^2}
	\left[
	\left(
	g_\PP^2 - \frac
	{ 2 \ii V g_\PP g_\MM
	}{
		\ii ( \omega_{\MM,\qq} - \omega_\PP ) + (\gamma_{\MM,\qq} - \gamma_\PP)/2
	} 
	\right) 
	\ee^{-\ii(\omega_\PP+\gamma_\PP/2)t}
	+
	g_\MM^2 \ee^{-\ii(\omega_{\MM,\qq}+\gamma_{\MM,\qq}/2)t}
	\right]
	\right\}.
\end{equation}

We approximate the magnon density of states, $\mathrm{DOS}(\omega)$, by two distinct modes as
\begin{equation} \label{DOS}
	\mathrm{DOS}(\omega) = \alpha\delta(\omega-\omega_{\MM_\LL})+(1-\alpha)\delta(\omega-\omega_{\MM_\HH}),    
\end{equation}
replacing 
\begin{equation}
	\frac{2\pi}{S\sigma^2}\sum_\qq
	\ee^{-\qq^2/2\sigma^2}
	\ldots
	=
	\int\limits_0^\infty \mathrm{DOS}(\omega) \ldots \dd\omega.
\end{equation}
We consider $\omega_{\MM_\LL} < \omega_{\MM_\HH}$ and $\alpha > 0$.
The corresponding dissipation rates of the magnons are $\gamma_{\MM_\LL}$ and $\gamma_{\MM_\HH}$.
In this case,
\begin{multline} \label{reflectivity}
	\frac{\Delta R(t)}{R_0}
	=
	I_0 \Im
	\biggl[
	\left(
	g_\PP^2
	- 
	\frac
	{ 2 \ii V g_\PP g_\MM \alpha
	}{
		\ii ( \omega_{\MM_\LL} - \omega_\PP ) + (\gamma_{\MM_\LL} - \gamma_\PP)/2
	} 
	-
	\frac
	{ 2 \ii V g_\PP g_\MM (1-\alpha)
	}{
		\ii ( \omega_{\MM_\HH} - \omega_\PP ) + (\gamma_{\MM_\HH} - \gamma_\PP)/2
	} 
	\right) 
	\ee^{-\ii(\omega_\PP+\gamma_\PP/2)t}
	\\
	+
	\alpha g_\MM^2 \ee^{-\ii(\omega_{\MM_\LL}+\gamma_{\MM_\LL}/2)t}
	+
	(1-\alpha)g_\MM^2 \ee^{-\ii(\omega_{\MM_\HH}+\gamma_{\MM_\HH}/2)t}
	\biggr]
\end{multline}

The fast Fourier transformation (FFT) of the signal $\Delta R(t)/R_0$ between times $t_1$ and $t_2$ corresponds to the Fourier transformation (FT)
\begin{equation} \label{DFT}
	S_\mathrm{FT}(f) 
	=
	\abs{\frac{1}{\Delta t}\int\limits_{t_1}^{t_2} \frac{\Delta R(t)}{R_0} \ee^{\ii 2\pi f t} \dd{t}},
\end{equation}
where the frequency $f$ is measured in \unit{Hz}, and $\Delta t$ is the time step corresponding to the FFT of the experimental data.

When comparing $S_\mathrm{FT}(f)$ and $S_\mathrm{FFT}(f)$, we consider only the following discrete frequencies: $f_n = -f_\mathrm{max}/2 + f_\mathrm{max}(n-1)/N$, where $f_\mathrm{max} = \Delta t^{-1}$, $N$ is the number of time steps between $t_1$ and $t_2$, given by $N = 1 + (t_2 - t_1)/\Delta t$.

From \cref{reflectivity,DFT}, we obtain the final result
\begin{multline} \label{analytic spectrum}
	S_\mathrm{FT}(f) 
	=
	\\
	\frac{1}{\Delta t}
	I_0 
	\Biggl\lvert\Biggl[
	\left(
	g_\PP^2
	- 
	\frac
	{ 2 \ii V g_\PP g_\MM \alpha
	}{
		\ii ( \omega_{\MM_\LL} - \omega_\PP ) + (\gamma_{\MM_\LL} - \gamma_\PP)/2
	} 
	-
	\frac
	{ 2 \ii V g_\PP g_\MM (1-\alpha)
	}{
		\ii ( \omega_{\MM_\HH} - \omega_\PP ) + (\gamma_{\MM_\HH} - \gamma_\PP)/2
	} 
	\right)
	\frac
	{
		ie^{\ii (2\pi f - \omega_\PP + \ii \gamma_\PP/2)t_1} - 
		ie^{\ii (2\pi f - \omega_\PP + \ii \gamma_\PP/2)t_2} 
	}{
		2\pi f - \omega_\PP + \ii \gamma_\PP/2
	}
	\\
	+
	\alpha g_\MM^2
	\frac
	{
		ie^{\ii (2\pi f - \omega_{\MM_\LL} + \ii \gamma_{\MM_\LL}/2)t_1} - 
		ie^{\ii (2\pi f - \omega_{\MM_\LL} + \ii \gamma_{\MM_\LL}/2)t_2} 
	}{
		2\pi f - \omega_{\MM_\LL} + \ii \gamma_{\MM_\LL}/2
	} 
	+
	(1-\alpha)g_\MM^2 
	\frac
	{
		ie^{\ii (2\pi f - \omega_{\MM_\HH} + \ii \gamma_{\MM_\HH}/2)t_1} - 
		ie^{\ii (2\pi f - \omega_{\MM_\HH} + \ii \gamma_{\MM_\HH}/2)t_2} 
	}{
		2\pi f - \omega_{\MM_\HH} + \ii \gamma_{\MM_\HH}/2
	} 
	\Biggr]\Biggr\rvert.
\end{multline}

\subsection{Simulated spectra and fitting parameters}
\begin{figure}
	\centering
	\includegraphics[width=0.55\linewidth]{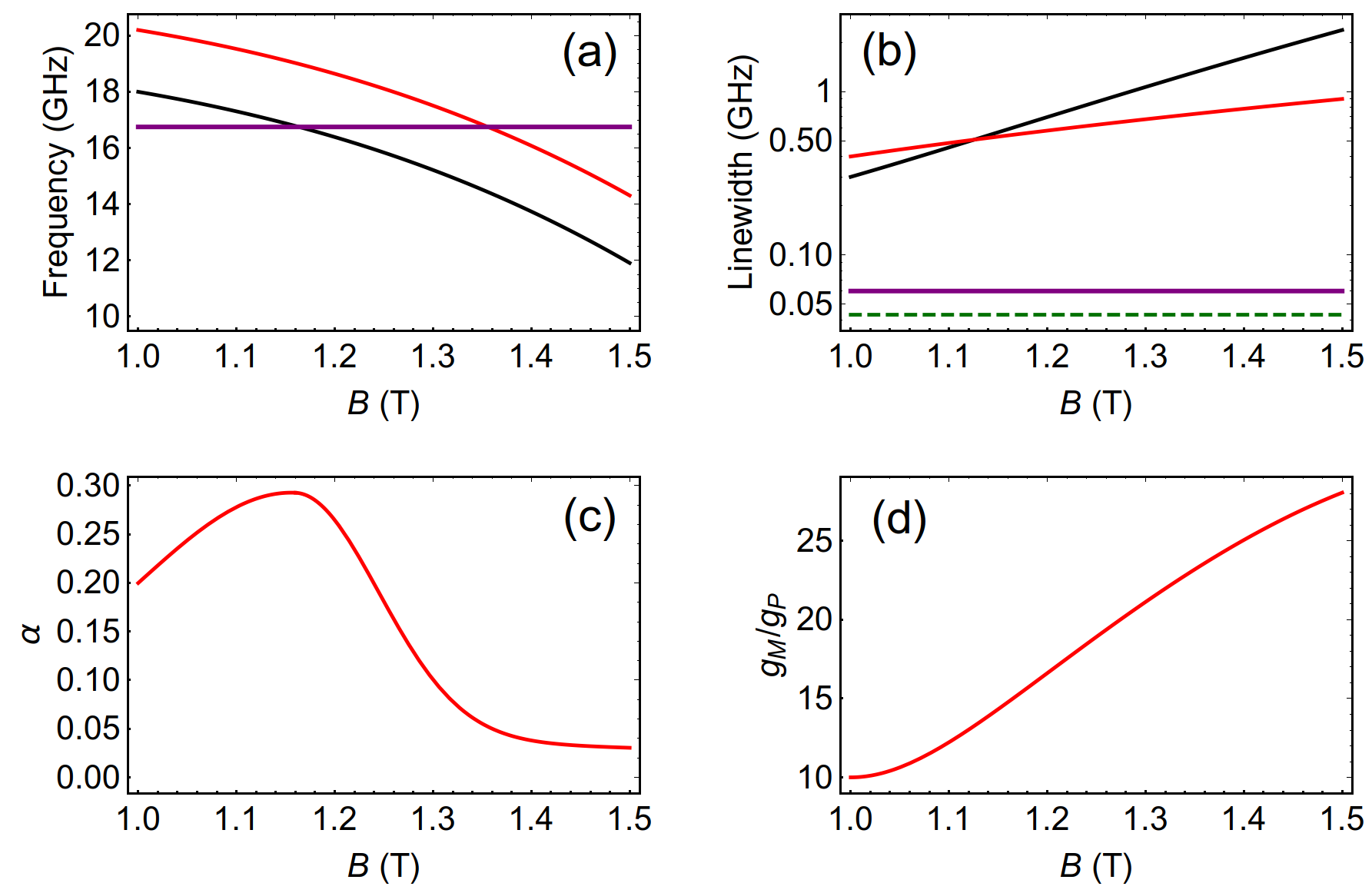}
	\caption{
		Fitted parameters.
		Parameters of the system obtained from the fit of the FFT spectra with \cref{DFT}.
		(a)~Frequencies $\omega_{\mathrm{M}_\mathrm{H}}/2\pi$ (red line), $\omega_{\mathrm{M}_\mathrm{L}}/2\pi$ (black curve), $\omega_\PP/2\pi$ (purple line).
		(b)~Linewidth $\gamma_{\mathrm{M}_\mathrm{H}}/2\pi$ (red line), $\gamma_{\mathrm{M}_\mathrm{L}}/2\pi$ (black line), $\gamma_\PP/2\pi$ (purple line), and $\abs{V/2\pi}$ (green dashed line).
		We note $V/2\pi = \SI{-43}{\MHz}$.
		(c)~Parameter $\alpha$ defined in \cref{DOS}.
		(d)~Ratio between interaction constants between polaritons and magnons $g_\MM$ and between polaritons and phonons $g_\PP$.
		\label{fig:Parameters}
	}
\end{figure}

The developed theory and \cref{analytic spectrum} allow us to obtain the analytical solution for the spectral response of the system plotted together with the experimental results in 
\cref{main-fig:4}(b--g) of the main text. To fit the FFT of the experimental data with the analytical expression~\eqref{analytic spectrum}, we follow the next 3 steps:

\begin{itemize}
	
	\item	{\it Step 1}:
	We fit the frequencies and linewidths of the two magnon branches ($\omega_{\mathrm{M}_\mathrm{H}}$, $\omega_{\mathrm{M}_\mathrm{L}}$, $\gamma_{\mathrm{M}_\mathrm{H}}$, $\gamma_{\mathrm{M}_\mathrm{L}}$), the parameter $\alpha$, and the interaction constant $g_\MM$ up to a factor $C$ from the FFT spectra in the time window \qtyrange{0.1}{1.9}{\ns}, shown in \cref{main-fig:4}(b) of the main text. The result of this fitting step is shown in \cref{fig:Parameters}(a--c).
	
	\item	{\it Step 2}:
	We fit the frequency and the linewidth of the phonons ($\omega_\PP$ and $\gamma_\PP$) as well as the interaction constant $g_\PP$ up to a factor $C$ (the same as in step 1) from the FFT spectra in the time window \qtyrange{3.8}{11.3}{\ns} at a magnetic field of \SI{1}{\T}, shown in \cref{main-fig:4}(d).
	The result of this fitting is shown in \cref{fig:Parameters}(a, b).
	We assume that the parameters of the phonons are independent of the applied magnetic field, which is supported by the out-of-resonance data where the phonon frequency is blue-shifted from the magnons.
	In this step, we also find $g_\MM/g_\PP$, shown in \cref{fig:Parameters}(d).
	
	\item	{\it Step 3}:
	Finally, we fit the interaction constant $V$ between phonons and magnons by considering the FFT spectra in the time window \qtyrange{3.8}{11.3}{\ns} at the full range of the magnetic field.
	This is the only independently fitted parameter in our model.
	From this fit, we obtain the value $V/2\pi = \SI{-43}{\MHz}$, as shown in \cref{fig:Parameters}(b).
	
\end{itemize}

Importantly, the negative sign of the interaction constant between phonons and magnons, at $g_\PP>0$ and $g_\MM>0$, follows from the fact that the amplitude of the phonons enhances when its frequency is lower than the magnon frequency.

\clearpage
\FloatBarrier
\section{Additional Data}

\begin{figure}[ht]
	\includegraphics{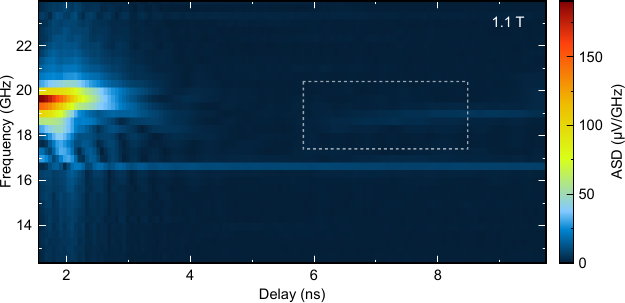}%
	\caption{\label{fig:Spectrogram}%
		Windowed Fourier transform for one of the measured pump--probe traces underlying \cref{main-fig:4}(b--d) at an applied external magnetic field of \SI{1.1}{\T} using a \SI{3}{\ns} long rectangular window. The delay axis marks the center of the Fourier window. The strong signature at \SI{19.5}{\GHz} visible for short delays are the high-momentum contributions of the magnon pulse, which quickly escape the probe region. The narrow line at \SI{16.8}{\GHz} is the fundamental breathing-type phonon mode of the suspended CrSBr flake. For long delays the low-momentum, localized magnon contributions show up as a second narrow line (dashed box). Their frequency notably blue-shifts with delay likely due to thermal effects, leading to a perceived lower $Q$-factor of the magnon spectrum in \cref{main-fig:4}(d).}%
\end{figure}

\FloatBarrier

\bibliography{local}